\documentclass{ws-ijmpd}

             \def\gray{$\gamma$-ray}
\def\grays{$\gamma$-rays}

 \def\epr{e-print astro-ph/}

\begin{document}

\markboth{F.W.   Stecker} {Testing  Relativity at  High  Energies with
Spaceborne Detectors}

%
\catchline{}{}{}{}{}
%

\title{TESTING   RELATIVITY   AT   HIGH  ENERGIES   USING   SPACEBORNE
DETECTORS\\
}

\author{\footnotesize F.W. STECKER}


\address{NASA Goddard Space Flight Center\\ Greenbelt, MD, USA}

\maketitle


\begin{abstract}

The {\it Gamma-Ray Large Area Space Telescope (GLAST)}, to be launched
in the fall of 2007, will measure the spectra of distant extragalactic
sources of high energy \grays, particularly active galactic nuclei and
\gray\  bursts.   {\it GLAST}  can  look  for energy-dependent  \gray\
propagation  effects  from  such   sources  as  a  signal  of  Lorentz
invariance  violation.  These sources  should  also  exhibit the  high
energy   cutoffs  predicted   to  be   the  result   of  intergalactic
annihilation interactions with low  energy photons having a flux level
as   determined   by    various   astronomical   observations.    Such
annihilations  result  in electron-positron  pair  production above  a
threshold energy given by $2m_{e}$  in the center of momentum frame of
the  system, assuming  Lorentz  invariance. If  Lorentz invariance  is
violated (LIV),  this threshold can be  significantly raised, changing
the  predicted absorption  turnover in  the observed  spectrum  of the
sources.   Stecker and  Glashow  have shown  that  the existence  such
absorption  features  in the  spectra  of  extragalactic sources  puts
constraints on LIV.  Such  constraints have important implications for
some  quantum  gravity  and   large  extra  dimension  models.  Future
spaceborne  detectors dedicated to  measuring \gray\  polarization can
look for  birefringence effects as  a possible signal of  loop quantum
gravity.

As  shown by Coleman  and Glashow,  a much  smaller amount  of Lorentz
invariance   violation  has   potential   implications  for   possibly
supressing  the   ``GZK  cutoff''  predicted  to  be   caused  by  the
interactions of  cosmic rays having multi-Joule  energies with photons
of  the 2.7  K  cosmic background  radiation  in intergalactic  space.
Owing  to the  rarity  of  such ultrahigh  energy  cosmic rays,  their
spectra are  best studied by  a UV-sensitive satellite  detector which
looks down  on a large volume  of the Earth's atmosphere  to study the
nitrogen fluorescence  tracks of  giant air-showers produced  by these
ultrahigh  energy  cosmic rays.   We  discuss  here  in particular,  a
two-satellite mission called {\it OWL},  which would be suited to make
such studies.

\end{abstract}

\section{Introduction}

The theory of  relativity is one of the  fundamental pillars of modern
physics.  However, because  of  the problems  associated with  merging
relativity with quantum theory, it  has long been felt that relativity
may  have to  be modified  in some  way.  It  has been  suggested that
relativity,  {\it  i.e.}  Lorentz  invariance  (LI), may  be  only  an
approximate   symmetry  of  nature\cite{sa72}.    There  has   been  a
particular interest in the  possibility that a breakdown of relativity
may be associated  with the Planck scale of  $M_{QG} \sim 10^{19}$ GeV
where  quantum   effects  are   expected  to  become   significant  in
gravitational  theory.  Although no  true  quantum  theory of  gravity
exists, it  was independently  proposed that LI  might be  violated in
such  a   theory  with  astrophysical   consequences\cite{ac98}  being
manifested at an  energy scale $<< M_{QG}$. The  subject of this paper
is the  potential use  of observations of  high energy  phenomena from
satellite detectors  to search for  the possible breakdown  of Lorentz
invariance.

\section{An LIV Formalism}

A  simple  formulation  for  breaking   LI  by  a  small  first  order
perturbation  in  the  electromagnetic  Lagrangian which  leads  to  a
renormalizable   treatment    has   been   given    by   Coleman   and
Glashow\cite{cg99}.   The small  perturbative  noninvariant terms  are
both  rotationally  and   translationally  invariant  in  a  preferred
reference frame  which one  can assume  to be the  frame in  which the
cosmic background  radiation is isotropic. These terms  are also taken
to   be  invariant  under   $SU(3)\otimes  SU(2)\otimes   U(1)$  gauge
transformations in the standard model.

With this  form of  LI violation (LIV),  different particles  can have
differing maximum  attainable velocities (MAVs) and these  MAVs can be
different from  $c$.  Using the formalism of  Ref.  \refcite{cg99}, we
denote the MAV of a particle  of type $i$ by $c_{i}$, a quantity which
is  not necessarily  equal to  $c \equiv  1$, the  low energy  {\it in
vacua\/} velocity of light.  We further define the difference $c_{i} -
c_{j} \equiv \delta_{ij}$.  These  definitions will be used to discuss
the  physics  implications  of  cosmic  ray  and  cosmic  $\gamma$-ray
observations\cite{sg01,st03,st05}.

In general then, $c_e \ne c_\gamma$. The physical consequences of such
a violation of  LI depend on the sign of  the difference between these
two MAVs. Defining

\begin{equation}
c_{e} \equiv c_{\gamma}(1 + \delta) ~ , ~ ~~~0< |\delta| \ll 1\;,
\end{equation}

\noindent
one  can consider the  two cases  of positive  and negative  values of
$\delta$ separately\cite{cg99,sg01}.

{\it Case I:} If $c_e<c_\gamma$ ($\delta  < 0$), the decay of a photon
into an  electron-positron pair  is kinematically allowed  for photons
with energies exceeding

\begin{equation}
E_{\rm max}= m_e\,\sqrt{2/|\delta|}\;.
\end{equation}

\noindent
The  decay would  take place  rapidly, so  that photons  with energies
exceeding $E_{\rm max}$ could not be observed either in the laboratory
or as cosmic rays. From the  fact that photons have been observed with
energies $E_{\gamma} \ge$ 50~TeV from the Crab nebula, one deduces for
this case  that $E_{\rm max}\ge  50\;$TeV, or that -$\delta  < 2\times
10^{-16}$.

{\it Case  II:} For this possibility, where  $c_e>c_\gamma$ ($\delta >
 0$), electrons  become superluminal if their  energies exceed $E_{\rm
 max}/2$.  Electrons  traveling faster than  light will emit  light at
 all frequencies by a process  of `vacuum \v Cerenkov radiation.' This
 process  occurs  rapidly,  so  that  superluminal  electron  energies
 quickly  approach $E_{\rm max}/2$.   However, because  electrons have
 been seen in the cosmic  radiation with energies up to $\sim\,$2~TeV,
 it  follows that $E_{\rm  max} \ge  2$~TeV, which  leads to  an upper
 limit on  $\delta$ for  this case of  $3 \times 10^{-14}$.  Note that
 this limit is two orders  of magnitude weaker than the limit obtained
 for  Case I.   However, this  limit can  be considerably  improved by
 considering constraints obtained from  studying the \gray\ spectra of
 active galaxies\cite{sg01}.

\section{Extragalactic Gamma-ray Constraints on LIV}

\begin{figure}[h]
\begin{center}
\psfig{figure=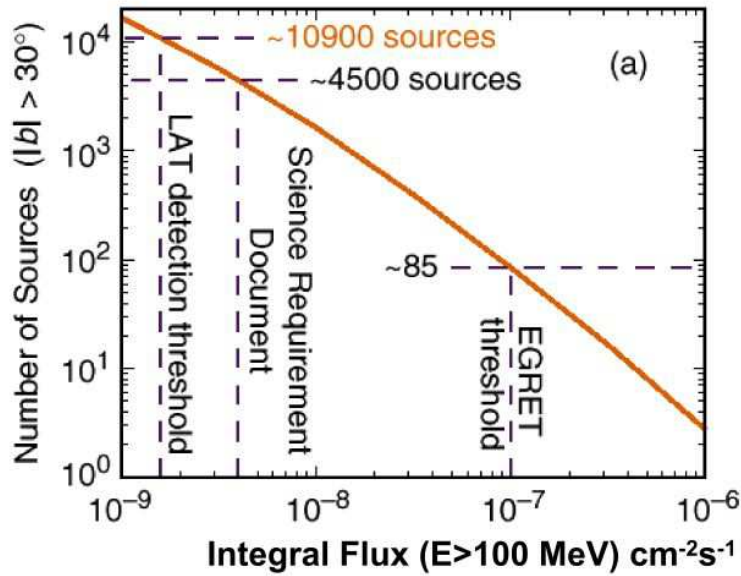,height=8cm}
\end{center}
\hspace{1.cm}
\caption{The  number  of  \gray\  emitting  active  galaxies  at  high
galactic latitudes (galactic latitude  $ |b| > 30 ^{\circ}$) predicted
to be seen  by the {\it GLAST LAT  (Large Area Telescope)} instrument.
The approximate number of sources detected by the previous {\it EGRET}
instrument  on   the  {\it  Cosmic  Gamma-Ray   Observatory}  is  also
shown. The curve  shows the predicted integral source  count {\it vs.}
threshold flux.\protect\cite{ss96}.}
\label{counts}
\end{figure}

\begin{figure*}[ht]
\begin{center}
\epsfxsize=14cm \epsfbox{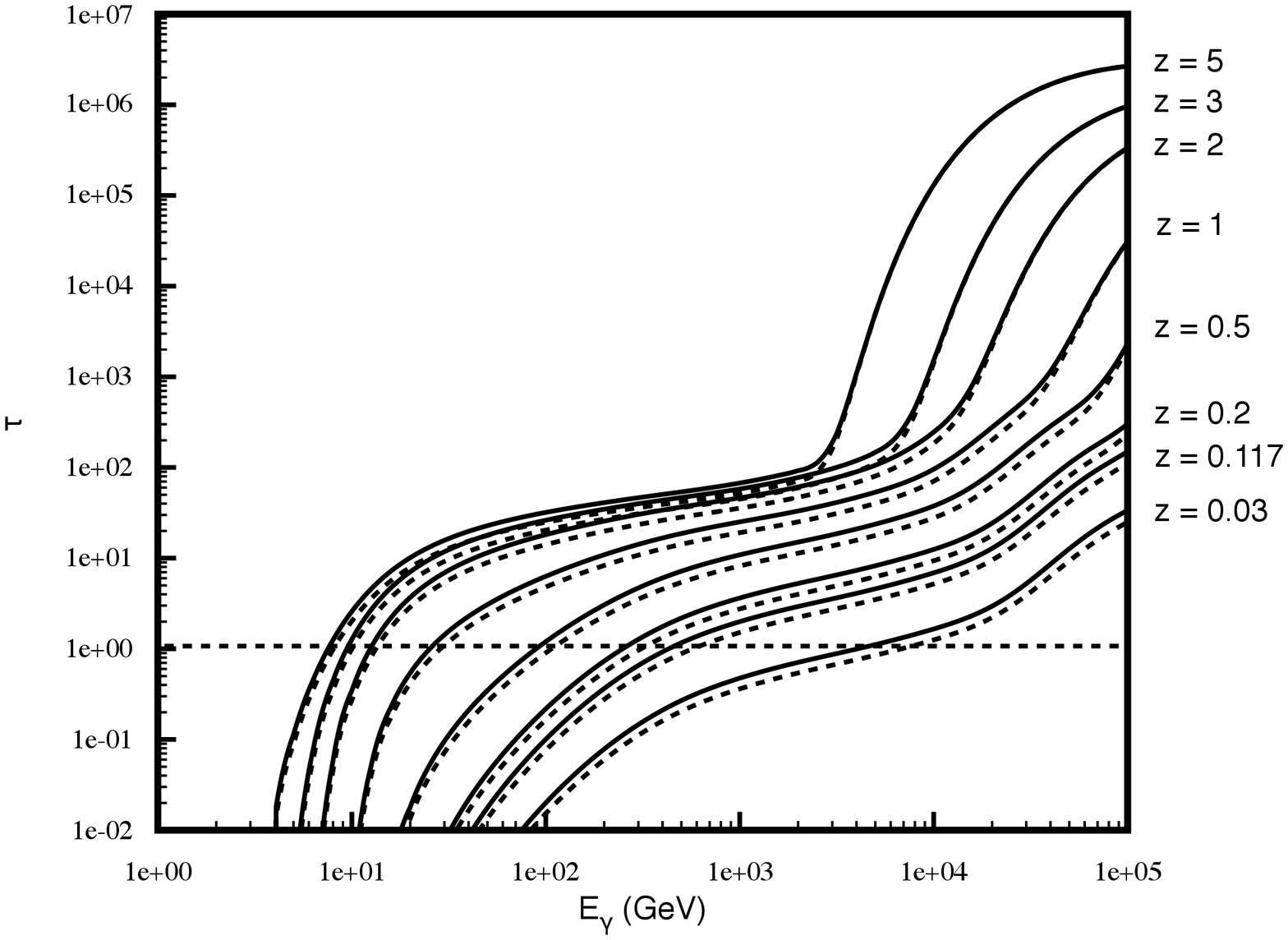}
\end{center}
\caption{The   optical  depth   of  the   universe  to   \grays\  from
interactions with  photons of  the intergalactic background  light and
the 2.7 K  cosmic background radiation for \grays\  having energies up
to 100 TeV.  This is given for a family of redshifts from 0.03 to 5 as
indicated.   The solid  lines are  for the  fast evolution  model; the
dashed lines are for the baseline model.\protect\cite{sms06}.}
\label{taufam}
\end{figure*}

A constraint on $\delta$ for $\delta > 0$ follows from a change in the
threshold  energy for  the pair  production process  $\gamma  + \gamma
\rightarrow e^+ + e^-$.  This follows from the fact that the square of
the four-momentum is changed to give the threshold condition

\begin{equation}
2\epsilon   E_{\gamma}(1-cos\theta)~   -~  2E_{\gamma}^2\delta   ~\ge~
4m_{e}^2,
\end{equation}

\noindent where $\epsilon$ is the  energy of the low energy photon and
$\theta$ is the angle between the  two photons. The second term on the
left-hand-side  comes  from  the  fact  that  $c_{\gamma}  =  \partial
E_{\gamma}/\partial p_{\gamma}$.  It follows  that the condition for a
significant increase  in the energy  threshold for pair  production is
$E_{\gamma}\delta/2$  $ \ge$  $ m_{e}^2/E_{\gamma}$,  or equivalently,
$\delta \ge {2m_{e}^{2}/E_{\gamma}^{2}}$.  ~ The observed $\gamma$-ray
spectrum  of the  active  galaxy  Mkn 501  while  flaring extended  to
$E_{\gamma}  \ge  24$ TeV\cite{ah01}  and  exhibited  the high  energy
absorption  expected from  $\gamma$-ray annihilation  by extragalactic
pair-production     interactions    with     extragalactic    infrared
photons\cite{ds02,ko03}.  This has  led Stecker and Glashow\cite{sg01}
to  point  out  that  the  Mkn  501  spectrum  presents  evidence  for
pair-production with  no indication  of LIV up  to a photon  energy of
$\sim\,$20~TeV and  to thereby place a quantitative  constraint on LIV
given  by  $\delta  <  2m_{e}^{2}/E_{\gamma}^{2} \simeq  10^{-15}$,  a
factor  of 30 better  than that  given in  the previous  section. {\it
GLAST}  will  observe many  more  such  active  galaxies at  different
redshifts  as  shown  in  Figure \ref{counts}\cite{ss96}  and  thereby
further test  such constraints on  LIV by looking for  deviations from
predicted  absorption effects. Figure  \ref{taufam} shows  the optical
depth of the  universe to high energy \grays\  against pair production
interactions  for sources  at various  redshifts under  the assumption
that Lorentz invariance holds\cite{sms06}.

\section{Gamma-ray Constraints on Quantum Gravity and Extra Dimension Models}

As previously mentioned, LIV has  been proposed to be a consequence of
quantum gravity physics at  the Planck scale $M_{Planck} = \sqrt{\hbar
c/G}  \simeq  1.22  \times  10^{19}$ GeV\cite{ga95,al02}.   In  models
involving large  extra dimensions, the  energy scale at  which gravity
becomes  strong can  occur at  a scale,  $M_{QG} <<  M_{Planck}$, even
approaching a  TeV\cite{el01}.  In the most  commonly considered case,
the  usual  relativistic   dispersion  relations  between  energy  and
momentum of  the photon and the  electron are modified\cite{ac98,al02}
by  a term  of  order $p^3/M_{QG}$.\footnote{We  note  that there  are
variants of quantum gravity and  large extra dimension models which do
not  violate   Lorentz  invariance  and  for   which  the  constraints
considered here do not apply.  There are also variants for which there
are no cubic terms in  momentum, but rather much smaller quartic terms
of order $\sim {p{^4}/ M_{QG}^2}$.}

Generalizing the LIV parameter $\delta$ to an energy dependent form

\begin{equation}
\delta~  \equiv~ {\partial  E_{e}\over{\partial p_{e}}}~  -~ {\partial
E_{\gamma}         \over{\partial         p_{\gamma}}}~        \simeq~
{E_{\gamma}\over{M_{QG}}}~      -~{m_{e}^{2}\over{2E_{e}^{2}}}~     -~
{E_{e}\over{M_{QG}}} ,
\end{equation}

\noindent the threshold condition from pair production implies $M_{QG}
~\ge~  E_{\gamma}^3/8m_{e}^2.$   Since  pair  production   occurs  for
energies  of at  least 20  TeV, we  find a  constraint on  the quantum
gravity scale\cite{st03} $M_{QG} \ge 0.3 M_{Planck}$.  This constraint
contradicts the  predictions of  some proposed quantum  gravity models
involving large extra dimensions  and smaller effective Planck masses.
In  a variant  model  of Ref.  \refcite{el04},  the photon  dispersion
relation is changed, but not that  of the electrons.  In this case, we
find the even stronger  constraint $M_{QG} \ge 0.6 M_{Planck}$. Future
studies of the spectra of active galaxies can extend these constraints
on quantum gravity models.

\section{Energy Dependent Time Variability of GRB Spectra and Tests of 
Lorentz Invariance Violation}

One possible  manifestation of Lorentz  invariance violation, possibly
from Planck  scale physics produced  by quantum gravity effects,  is a
change in  the energy-momentum dispersion relation of  a free particle
or a photon  which may be of first  order in $E_{\gamma}/M_{QG}$ where
$M_{QG}$  is the  quantum gravity  scale,  usually assumed  to be  the
Planck  scale\cite{ac98,el01}.  In  a $\Lambda  CDM$  cosmology, where
present observational data indicate that $\Omega_{\Lambda} \simeq 0.7$
and  $\Omega_{m}   \simeq  0.3$,  the  resulting   difference  in  the
propagation times  of two photons having an  energy difference $\Delta
E_{\gamma}$ from a $\gamma$-ray burst (GRB) at a redshift $z$ will be

\begin{equation}
\Delta  t_{LIV}   =  H_{0}^{-1}  {{\Delta   E_{\gamma}}  \over  M_{QG}
}{\int_0^z}{{dz'}        \over        {\sqrt{\Omega_{\Lambda}        +
\Omega_{m}(1+z')^3}}}
\end{equation}

\noindent for  a photon dispersion of  the form $c_{\gamma}  = c(1 \pm
E_{\gamma}/M_{QG}$), with  $c$ being the usual low  energy velocity of
light\cite{el03,el06}. In  other words, $\delta$,  as defined earlier,
is given by $\pm E_{\gamma}/M_{QG}$. Data on GRB021206 for $E_{\gamma}
> 3$  MeV  implies   a  value  for  $M_{QG}  >   1.8  \times  10^{17}$
GeV\cite{sb04}. Data from GRB051221A have given a constraint $M_{QG} >
0.66 \times 10^{17}$ GeV\cite{ro06}.

The dispersion effect will be smaller if the dispersion relation has a
quadratic dependence on  $E_{\gamma}/M_{QG}$ as suggested by effective
field  theory considerations\cite{my03,ja04}.  This  will obviate  the
limits on $M_{QG}$ given above. The possible effect of extra dimension
models  on   \gray\  propagation  has  also  been   pointed  out  very
recently\cite{go06}.

\begin{figure}[h]
\begin{center}
\psfig{figure=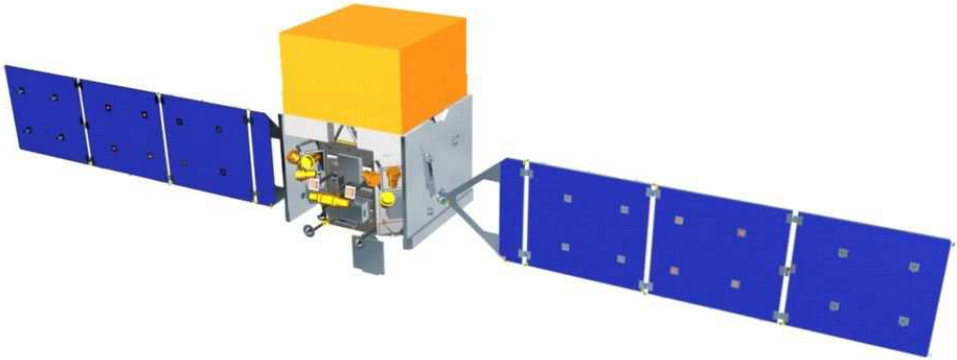,height=5cm}
\end{center}
\hspace{1.cm}
\caption{Schematic of the {\it GLAST} satellite deployed in orbit. The
{\it LAT}  is in the top (yellow)  area and the {\it  GBM} are located
directly below.}
\label{glast}
\end{figure}

The {\it  GLAST} satellite, (see Figure \ref{glast}),  with its \gray\
{\it Burst Monitors (GBM)} covering an  energy range from 10 keV to 25
MeV and its {\it Large  Area Telescope (LAT)} covering an energy range
from 20 MeV to $> 300$ GeV, can study both GRBs and flares from active
galactic nuclei over a large range of both energy and distance. So our
studies can be extended to {\it GLAST} observations of GRBs and blazar
flares after the expected {\it GLAST} launch in the fall of 2007.

\section{Looking for Birefringence Effects from Quantum Gravity}

A  possible model for  quantizing space-time  which has  been actively
investigated is  {\it loop quantum  gravity} (see the review  given in
Ref. \refcite{pe04} and references therein.) A signature of this model
is  that the  quantum nature  of space-time  can produce  an intrinsic
birefringence  effect.  This   is  because  electromagnetic  waves  of
opposite   circular  polarizations   will  propagate   with  different
velocities, which leads to a rotation of linear polarization direction
through the angle
\begin{equation} 
\theta(t)=\left[\omega_+(k)-\omega_-(k)\right]t/2=\xi               k^2
t/2M_{Planck}
\label{rotation}
\end{equation}
for a plane wave with wave-vector $k$\cite{ga99}.

Some astrophysical sources emit  highly polarized radiation. It can be
seen from equation (\ref{rotation}) that the rotation angle is reduced
by the  large value of the  Planck mass. However,  the small rotations
given  by equation (\ref{rotation})  can add  up over  astronomical or
cosmological  distances  to  erase  the  polarization  of  the  source
emission. Therefore, if  polarization is seen in a  distant source, it
puts constraints  on the  parameter $\xi$.  Observations  of polarized
radiation from distant sources can therefore be used to place an upper
bound on $\xi$.

Equation (\ref{rotation})  indicates that  the higher the  wave number
$|k|$,  the   stronger  the  rotation   effect  will  be.   Thus,  the
depolarizing effect  of space-time induced birefringence  will be most
pronounced in  the \gray\ energy range.  It can also be  seen that the
this effect  grows linearly with  propoagation time.  The  best secure
bound on  this effect,  $|\xi|\lesssim 2\times 10^{-4}$,  was obtained
using  the observed  10\%  polarization of  ultraviolet  light from  a
distant galaxy\cite{gk01}.

A  few  years  ago,  there  was  a  report  of  strong  linear  \gray\
polarization from  the \gray\ burst  GRB021206 observed from  the {\it
RHESSI} satellite\cite{cb03}.  The survival of  such polarization over
cosmological  distances would put  a much  stronger constraint  on the
value of the parameter $\xi$. The constraint arises from the fact that
if the angle of  polarization rotation (\ref{rotation}) were to differ
by more than $\pi/2$  over the 0.1 - 0.3 MeV energy  range and by more
than  $3\pi/2$ over the  0.1-0.5 MeV  energy range,  the instantaneous
polarization at the detector  would fluctuate sufficiently for the net
polarization of  the signal to  be suppressed well below  the observed
value.  The  difference in rotation angles for  wave-vectors $k_1$ and
$k_2$ is
\begin{equation}
 \Delta\theta=\xi (k_2^2-k_1^2) d/2M_{Planck},
  \label{diffrotation}
\end{equation}
replacing the time  $t$ by the distance from the  GRB to the detector,
denoted by $d$.

While the distance to GRB021206 is unknown, it is well known that most
cosmological bursts  have redshifts in the range  1-2 corresponding to
distances  of greater  than  a Gpc.  Using  the distance  distribution
derived in Ref.~\refcite{DLRG} one can conservatively take the minimum
distance  to this burst  as 0.5  Gpc, corresponding  to a  redshift of
$\sim 0.1$. This yields the constraint
\begin{equation}
|\xi|<5.0\times10^{-15}/d_{0.5}.
\end{equation}
where  $d_{0.5}$  is  the  distance  to  the burst  in  units  of  0.5
Gpc\cite{ja04}.   However, the  polarization  measurement reported  in
Ref.     \refcite{cb03}    has     been     questioned    in     other
analyses\cite{rf04,wi04} and so remains controversial.

It should  be noted  that the {\it  RHESSI} satellite detctor  was not
designed specifically to  measure \gray\ polarization. Detectors which
are  dedicated to polarization  measurements in  the X-ray  and \gray\
energy range and which can be flown in space to study the polarization
from     distant     astronomical     sources    are     now     being
designed\cite{mi05,pr05}.  We note that  linear polarization  in X-ray
flares from GRBs has been predicted\cite{fa05}.

A further discussion of astrophyical constraints on LIV may be found in 
Ref.\refcite{ja04}.

\section{LIV and the Ultrahigh Energy Cosmic Ray Spectrum}

The flux of ultrahigh energy  nucleons is expected to be attenuated by
photomeson  producing interactions  of these  hadrons with  the cosmic
microwave background radiation (CBR). This  predicted effect is now known as
the  ``GZK effect''\cite{gr66,za66}.  The  mean-free-path for
this attenuation effect  is less than 100 Mpc  for cosmic ray nucleons
of energy greater than 100  EeV\cite{st68}.

Coleman and  Glashow\cite{cg99} have  shown that for  interactions of
protons with CBR photons of energy $\epsilon$ and temperature $T_{CBR}
= 2.73  K$, pion production  is kinematically forbidden and  thus {\it
photomeson interactions are turned off} if

\begin{equation}
\delta_{p\pi} > 5 \times 10^{-24}(\epsilon/T_{CBR})^2.
\end{equation}

\begin{figure}
\vspace{-1.5cm} \centerline{\psfig{figure=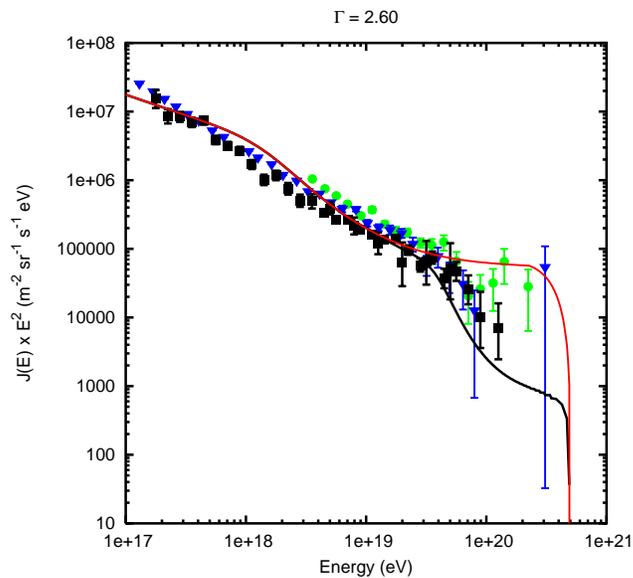,height=12cm}}
\vspace{-1.5cm}
\caption{Predicted  spectra  for an  $E^{-2.6}$  source spectrum  with
redshift evolution and $E_{max}$ = 500 EeV, shown with pair-production
losses included and photomeson  losses both included (black curve) and
turned off (lighter (red) curve).  The curves are shown with ultrahigh
energy cosmic ray spectral data from {\it Fly's Eye} (triangles), {\it
AGASA\protect\cite{ta98}} (circles) and {\it HiRes}\protect\cite{hr04}
monocular data (squares)\protect\cite{st05}.}
\label{spec}
\end{figure}

Thus, given even  a very small amount of  LIV, photomeson interactions
of  ultrahigh energy cosmic  rays (UHECR)  with the  cosmic background
radiation  can  be  turned  off.  Such  a  violation of  Lorentz
invariance  might  be  produced  by Planck  scale  effects\cite{al00,ap03}.

Some ``trans-GZK'' hadronic showers  with energies above the predicted
``cutoff energy'' (usually considered to be 100 EeV) have been observed
by both  scintillator and fluorescence detectors,  particularly by the
scintillator  array  {\it  AGASA}  group at  Akeno,  Japan\cite{ta98},
possibly in  contradiction to the expected  attenuation effect.  While
there is  less evidence for such interesting  events from fluorescence
detectors (see  Figure \ref{spec}), we  note that the {\it  Fly's Eye}
fluorescence   detector  reported   the   detection  of   a  320   EeV
event\cite{bi95},  an energy  which is  a factor  of $\sim$5  above  the GZK
cutoff energy.   The subject of  UHECRs having trans-GZK  energies has
not as yet been settled experimentally, even by the {\it Pierre Auger}
ground-based detector array\cite{au06}.

If  Lorentz   invariance
violation is the explanation for a possibly missing GZK effect, indicated in
the  {\it  AGASA} data  but  not the  {\it  HiRes}  data (see  Figure
\ref{spec}\cite{st05}), one  can also  look  for the  absence of  a
``pileup''  spectral  feature  and   for  the  absence  of the neutrinos 
which should be produced by the GZK effect. The detection of ultrahigh
energy nucleons and neutrinos at sufficiently high energies and with
excellent event statistics can best be done from space. This possibility 
will be discussed in the next section.

\section{The {\it OWL} Satellite Detectors}

The  {\it  OWL}  (Orbiting  Wide-field  Light-collectors)  mission  is
designed to  obtain data on  ultrahigh energy cosmic rays  (UHECR) and
neutrinos in order to  tackle the fundamental problems associated with
their  origin\cite{Sk04}.  The  {\it  OWL}  mission  is designed  to
provide  the  event statistics  and  extended  energy  range that  are
crucial  to addressing these  issues.  To  accomplish this,  {\it OWL}
makes use of  the Earth's atmosphere as a  huge ``calorimeter" to make
stereoscopic measurements of  the atmospheric UV fluorescence produced
by air shower particles. This  is the most accurate technique that has
been  developed  for  measuring  the energy,  arrival  direction,  and
interaction characteristics of UHECR\cite{Str98}.  To this end, {\it
OWL} will  consist of a pair of  satellites placed in tandem  in a low
inclination,  medium altitude  orbit.  The {\it  OWL} telescopes  will
point  down at  the Earth  and  will together  point at  a section  of
atmosphere about the size of the  state of Texas ($\sim 6 \times 10^5$
km$^2$),  obtaining a  much  greater sensitivity  than present  ground
based detectors. The ability of {\it OWL} to detect cosmic rays, in
units of km$^2$ sr, is called the {\it aperture}. The instantaneous aperture
at the highest  energies is $\sim 2 \times 10^6$  km$^2$ sr.  
The effective  aperture, reduced  by the  effects of  the moon,
man-made light,  and clouds, will  be conservatively $\sim  0.9 \times
10^5$ km$^2$ sr.   For each year of operation, {\it  OWL} will have 90
times the  aperture of  the ground based  {\it HiRes} detector  and 13
times the aperture of the {\it Pierre Auger} detector array (130 times
its  most sensitive  ``hybrid"  mode). The  {\it  OWL} detectors  will
observe the UV fluorescence light  from the giant air showers produced
by UHECR  on the  dark side of  the Earth.   They will thus  produce a
stereoscopic picture  of the temporal  and spatial development  of the
showers.\footnote{ The technical details, as well as discussion of the
science, including  ultrahigh energy neutrino science  with {\it OWL},
can be found at {\tt http://owl.gsfc.nasa.gov}.}

\begin{figure}[h]
\begin{center}
\mbox{\psfig{figure=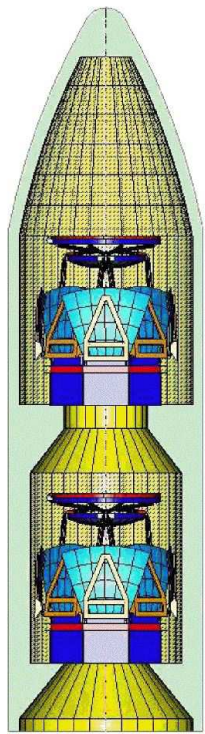,height=10cm}}
\end{center}
\hspace{1.cm}
\caption{Schematic of  the stowed {\it  OWL} satellites in  the launch
vehicle.}
\label{delta}
\end{figure}

\begin{figure}[h]
\begin{center}
\psfig{figure=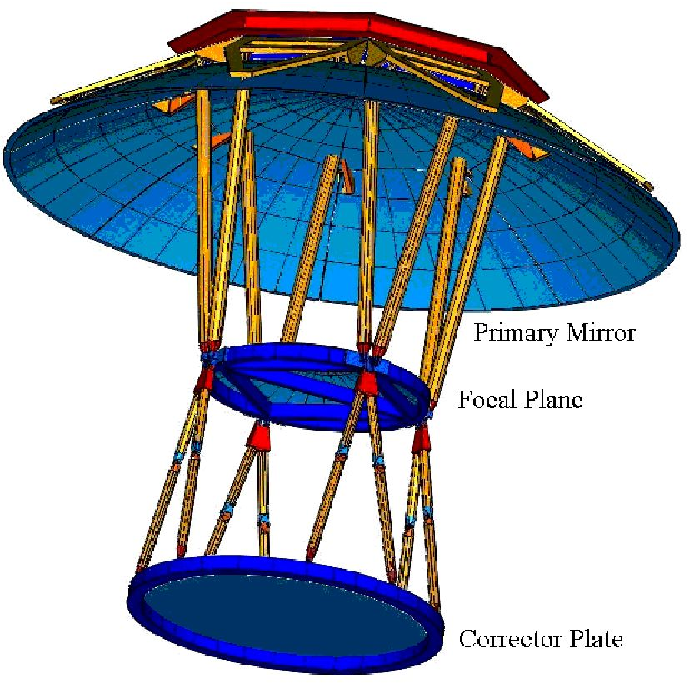,height=9cm}
\end{center}
\hspace{1.cm}
\caption{Schematic  of  the Schmidt  optics  that  form  an {\it  OWL}
``eye''  in the  deployed  configuration.  The  spacecraft bus,  light
shield, and shutter are not shown.}
\label{Schmidt}
\end{figure}

Following a stacked  dual launch on a Delta rocket  as shown in Figure
\ref{delta},  the  two  satellites   will  fly  in  formation at an
altitude of 1000 km and with  a
separation of  10 to  20 km for  about 3  months to search  for upward
going showers  from $\nu_{\tau}$'s propagating through  the Earth. The
spacecraft  will then  separate  to 600  km  for $\sim$  2.5 years  to
measure  the high-energy end  of the  UHECR spectrum.   Following this
period, the altitude is reduced to 600 km and the separation to 500 km
in order to measure the cosmic ray flux closer to 10 EeV.

With the  fluorescence technique, a fast, highly  pixelized camera (or
``eye") is used  to resolve both the spatial  and temporal development
of the shower.  This detailed information provides a powerful tool for
determining  the nature  of the  primary particle.   The  UV emission,
principally in  the 300 to 400  nm range, is isotropic  and the camera
can view the shower from  any direction, except almost directly toward
the camera. In the exceptional  case, the camera may still be utilized
as  a Cherenkov  detector. Thus,  a single  camera can  view particles
incident on the Earth from a hemisphere of sky.

In monocular operation, precision measurements of the arrival times of
UV photons  from different parts of  the shower track must  be used to
partly resolve  spatial ambiguities. The angle of  the shower relative
to    the   viewing   plane    is   resolvable    using   differential
timing. Resolving distance, however,  requires that the pixel crossing
time  be measured  to  an  accuracy that  is  virtually impossible  to
achieve  in  a  real  instrument  at  orbit  altitudes.   Stereoscopic
observation resolves both of these ambiguities. In stereo, fast timing
provides supplementary  information to reduce  systematics and improve
the resolution of the arrival direction of the UHECR. By using stereo,
differences in  atmospheric absorption or  scattering of the  UV light
can  be   determined.   The  results  obtained  by   the  {\it  HiRes}
collaboration  viewing the  same  shower in  both  modes have  clearly
demonstrated the desirability of stereo viewing.

The  light collector  will use  a Schmidt  camera design  as  shown in
Figure  \ref{Schmidt}. The  Schmidt  corrector has  a spherical  front
surface and an  aspheric back surface, while the  primary mirror has a
slight aspheric figure.  The focal  plane is a spherical surface tiled
with  flat detector  elements.  The  corrector is  slightly  domed for
strength. The primary is made of lightweight composite material with a
central  octagonal  section and  eight  petals  that  fold upward  for
launch. The  entire optical system  is covered by an  inflatable light
and  micrometeoroid shield  and is  closed out  by a  redundant shutter
system. The  shield will  be composed of  a multi-layer  material with
kevlar layers for strength.

Monte  Carlo  simulations of  the  physics  and  response of  orbiting
instruments  to the  UV air  fluorescence signals  are crucial  to the
development of {\it  OWL}. One such Monte Carlo  has been developed at
the NASA  Goddard Space  Flight Center {\cite{Kr01}}.   The simulation
employs  a  hadronic event  generator  that  includes  effects due  to
fluctuation  in  the shower  starting  point  and shower  development,
charged pion decay, neutral pion interactions, and the LPM (Landau-
Migdal-Pomeranchuk) effect.

\begin{figure}
\centerline{\psfig{figure=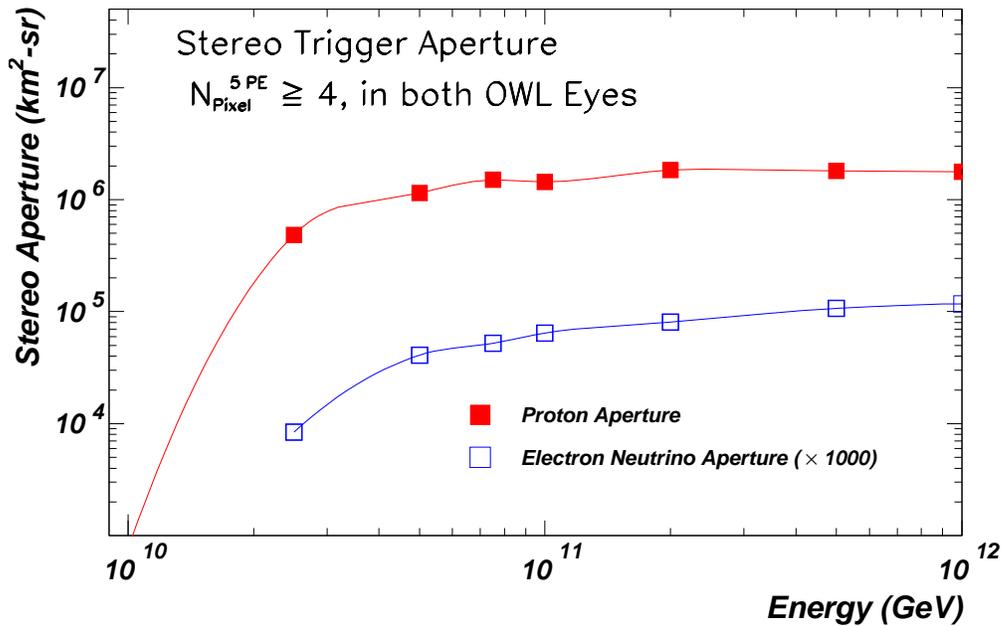,height=10cm}}
\caption{Instantaneous   aperture    for   proton-induced   and   deep
$\nu_{e}$-induced giant air-showers as a function of energy.}
\label{aper}
\end{figure}

The number  of events  detected by OWL  for a  monoenergetic isotropic
flux of protons and $\nu_{e}$'s with a standard model cross section is
calculated by the Monte Carlo program, yielding the detection aperture
as   a   function   of    energy,   simulated   trigger,   and   orbit
parameters. Figure \ref{aper} shows  the resultant proton and neutrino
aperture for  an altitude of 1000 km  and a separation of  500 km. The
asymptotic  instantaneous  proton aperture  is  $\sim  2 \times  10^6$
km$^2$sr.   The   $\nu_{e}$   aperture  determination   includes   the
requirement  that  the observed  starting  point  of  the air  shower,
$X_{start} \ge 1500$ g cm$^{-2}$ in slant depth.

\section*{Acknowledgments}

Part of this work was supported by NASA grant ATP03-0000-0057.

\end{document}